\documentclass[conference]{IEEEtran}

\usepackage{scalerel}
\usepackage{tikz}
\usetikzlibrary{svg.path}

\definecolor{orcidlogocol}{HTML}{A6CE39}
\tikzset{
  orcidlogo/.pic={
    \fill[orcidlogocol] svg{M256,128c0,70.7-57.3,128-128,128C57.3,256,0,198.7,0,128C0,57.3,57.3,0,128,0C198.7,0,256,57.3,256,128z};
    \fill[white] svg{M86.3,186.2H70.9V79.1h15.4v48.4V186.2z}
                 svg{M108.9,79.1h41.6c39.6,0,57,28.3,57,53.6c0,27.5-21.5,53.6-56.8,53.6h-41.8V79.1z M124.3,172.4h24.5c34.9,0,42.9-26.5,42.9-39.7c0-21.5-13.7-39.7-43.7-39.7h-23.7V172.4z}
                 svg{M88.7,56.8c0,5.5-4.5,10.1-10.1,10.1c-5.6,0-10.1-4.6-10.1-10.1c0-5.6,4.5-10.1,10.1-10.1C84.2,46.7,88.7,51.3,88.7,56.8z};
  }
}

\newcommand\orcidicon[1]{\href{https://orcid.org/#1}{\mbox{\scalerel*{
\begin{tikzpicture}[yscale=-1,transform shape]
\pic{orcidlogo};
\end{tikzpicture}
}{|}}}}

\usepackage{comment}
\usepackage{float}
\usepackage{siunitx}
\usepackage{paralist}
\usepackage{multirow}
\usepackage{overpic}
\usepackage{graphicx}

\usepackage{amsmath}
\usepackage{array}
\usepackage{listings}
\usepackage{caption}
\usepackage{subcaption}
\usepackage{svg}

\usepackage[T1]{fontenc}
\usepackage{inconsolata}
\usepackage{listings}
\usepackage{xcolor}

\lstdefinestyle{cpp}{
    language=C++,
    basicstyle=\scriptsize\ttfamily,
    keywordstyle=\color{blue},
    commentstyle=\color{gray},
    stringstyle=\color{red},
    numberstyle=\tiny\color{gray},
    stepnumber=1,
    numbersep=4pt,
    showstringspaces=false,
    breaklines=true,
    frame=lines,
    tabsize=2,
    xleftmargin=.1in,
    xrightmargin=.1in,
    numbers=left,
}

\usepackage[sorting=none]{biblatex} 
\usepackage{comment}

\usepackage[colorlinks=true, linkcolor=blue, citecolor=blue, urlcolor=blue]{hyperref}

\DeclareRobustCommand{\ppmetric}{%
  \tikz[baseline=(P.base)]{
    \node[inner sep=0pt] (P) {
      \makebox[3.2pt][l]{\scalebox{-1}[1]{$ \mathrm{P} $}}%
      $ \mathrm{P} $
    };
  }%
}

\newcommand{\linebreakand}{%
  \end{@IEEEauthorhalign}
  \hfill\mbox{}\par
  \mbox{}\hfill\begin{@IEEEauthorhalign}
}

\begin{document}
%
\title{Towards Portability at Scale: A Cross-Architecture Performance Evaluation of a GPU-enabled Shallow Water Solver}
\author{
\IEEEauthorblockN{Johansell Villalobos\IEEEauthorrefmark{1} - jovillalobos@cenat.ac.cr \orcidicon{0009-0002-3398-0714}}
\IEEEauthorblockN{Daniel Caviedes-Voulli\`eme\IEEEauthorrefmark{2}\IEEEauthorrefmark{3} - d.caviedes.voullieme@fz-juelich.de \orcidicon{0000-0001-7871-7544}}
\IEEEauthorblockN{Silvio Rizzi\IEEEauthorrefmark{4} - srizzi@anl.gov \orcidicon{0000-0002-3804-2471}}
\IEEEauthorblockN{Esteban Meneses\IEEEauthorrefmark{1}\IEEEauthorrefmark{5} - emeneses@cenat.ac.cr \orcidicon{0000-0002-4307-6000}}
\vspace{0.1cm}
\IEEEauthorblockA{\IEEEauthorrefmark{1}National High Technology Center, San José, Costa Rica}
\IEEEauthorblockA{\IEEEauthorrefmark{2}Simulation and Data Lab. Terrestrial Systems, Jülich Supercomputing Center}
\IEEEauthorblockA{\IEEEauthorrefmark{3}Institute for Bio- and Geosciences: Agrosphere (IBG-3), Forschungzentrum Jülich, Germany}
\IEEEauthorblockA{\IEEEauthorrefmark{4}Argonne National Laboratory, Lemont, Illinois, US}
\IEEEauthorblockA{\IEEEauthorrefmark{5}, Costa Rica Technological Institute, Cartago, Costa Rica}
}

\maketitle
\begin{abstract}
   Current climate change has posed a grand challenge in the field of numerical modeling due to its complex, multiscale dynamics. In hydrological modeling, the increasing demand for high-resolution, real-time simulations has led to the adoption of GPU-accelerated platforms and performance portable programming frameworks such as Kokkos. In this work, we present a comprehensive performance study of the SERGHEI-SWE solver, a shallow water equations code, across four state-of-the-art heterogeneous HPC systems: Frontier (AMD MI250X), JUWELS Booster (NVIDIA A100), JEDI (NVIDIA H100), and Aurora (Intel Max 1550). We assess strong scaling up to 1024 GPUs and weak scaling upwards of 2048 GPUs, demonstrating consistent scalability with a speedup of 32 and an efficiency upwards of 90\% for most almost all the test range. Roofline analysis reveals that memory bandwidth is the dominant performance bottleneck, with key solver kernels residing in the memory-bound region. To evaluate performance portability, we apply both harmonic and arithmetic mean-based metrics while varying problem size. Results indicate that while SERGHEI-SWE achieves portability across devices with tuned problem sizes (<70\%), there is room for kernel optimization within the solver with more granular control of the architecture specifically by using Kokkos teams and architecture specific tunable parameters. These findings position SERGHEI-SWE as a robust, scalable, and portable simulation tool for large-scale geophysical applications under evolving HPC architectures with potential to enhance its performance.
\end{abstract}


\IEEEpeerreviewmaketitle

\section{Introduction}\label{sec:intro}

Flash flood forecasting has posed a grand challenge due to its phenomenological complexity, which arises from the combination of multiple factors, including highly variable precipitation patterns, complex terrain, land use changes, and nonlinear hydrological responses. Addressing this challenge requires Early Warning Systems (EWS) that are composed of high-resolution numerical models that can simulate water flow dynamics, integrate real-time sensor data and most importantly be computationally efficient to ensure appropriate lead times for decision making.

There have been efforts developed in this direction that are well established such as the European Flood Awareness System (EFAS), its global variant Global Flood Awareness System (GloFAS), and the Flooded Locations And Simulated Hydrographs (FLASH) Project. These systems integrate real-time meteorological and hydrological data with advanced numerical models to enhance flood forecasting capabilities. EFAS and GloFAS leverage large-scale deterministic atmospheric predictions and the LISFLOOD hydrological model operationally as well as the LISFLOOD-FP model, while FLASH utilizes high-resolution radar-based precipitation estimates to rapidly assess flash flood risks leveraging the CREST hydrological model. 

Despite having already established workflows for early warning and having success with these models, these lack high resolution in their forecasts providing results only up to 1km of resolution. In the context of flash flooding this can suffice to provide a general overview of the response of a  watershed. However, specific valleys and streams (usually ungauged) cannot be studied with the required detail. Indeed, at such 1km resolution, these valleys and streams are very poorly represented. Additionally, hydrological models are unable to represent wave propagation which is crucial to estimate time-to-flood. Instead, hydrodynamic models are used to properly forecast and achieve a better understanding of complex flooding events, as they consider the spatial and temporal variation of water depth and velocity. The accepted best choice for hydrodynamic models for this purpose are a system of 2D partial differential equations, better known as the Shallow Water Equations (SWE). 

Not until recently have SWE solvers been applied at very fine resolution and large domain sizes, primarly because of its high computational complexity. Advancements in numerical methods, hardware acceleration, and scalable parallel algorithms have made it possible to efficiently solve the SWE. More recently, the uptake of high performance computing (HPC) techniques, such as domain decomposition, GPU acceleration, and asynchronous communication strategies, has significantly reduced computational bottlenecks, enabling faster-than-realtime predictions at a high resolution and accuracy. In the era of exascale computing, an HPC enabled SWE model then becomes feasible and beneficial to implement in EWS given the information that we currently have and the finer predictions that it would provide.

EFAS has implemented this with the LISFLOOD-FP model that solves the SWE. However, their current implementation yields forecasts up to 100m in resolution. Although this is better than the 1km resolutions, flash floods usually act at the local level in which resolutions closer to $5-10 \mathrm{m}$ are more advantageous. 

There are several open-source SWE solvers that have been developed in the last decade that can provide results for resolutions around $1 \mathrm{m}$. This is favorable for flash flood prediction, however these results may be useless if computational walltime is excessive. As timely simulations are fundamental to ensure enough forecast lead time, the acceleration of computational methods is paramount. This can be accomplished using software that leverages the heterogeneous nature of current supercomputers.

Several SWE solvers have been developed to take advantage of graphics processing units (GPUs) as well as distributed programming techniques to accelerate the solution process. Most of these rely on the CUDA-Message Passing Interface (MPI) backend combination \cite{MORALESHERNANDEZ2021105034,Saleem2024} which sets them at a disadvantage when considering the current state of supercomputing platforms. 

Until June 2022, the TOP500 list predominantly featured supercomputers built around NVIDIA GPUs paired with CPUs. However, with the advent of systems like Frontier and Lumi, both leveraging AMD GPUs, or Aurora made with Intel GPUs, the high-performance computing landscape shifted. This transition further underscores the limitations of SWE solvers that depend on the CUDA-MPI backend, as they are not readily optimized for these emerging architectures.

Avoiding this vendor lock trap is crucial for the development of efficient, reliable, and transferable EWS, so that they can be deployed anywhere, regardless of the available hardware. This is, at the core, the issue of performance portability.

The capability of porting software to different computing architectures is particularly attractive in the context of projects like the exascale computing project (ECP) or the EuroHPC Joint Undertaking that invest in considerably large supercomputing resources to solve these specific problems \cite{Dubey2021Performance, EuroHPC_MASP_2024}.

In the case of the EuroHPC JU, one of its goals is to maintain a federated, hyper-connected supercomputing system built on platforms with completely different architectures. This diversity leads to the idea of leveraging these vast resources for use in EWS. Portable software then becomes an indispensable tool, allowing for the seamless deployment of EWS across heterogeneous computing environments. By abstracting away hardware-specific details, it mitigates the constraints of vendor lock-in while enhancing both the resilience and scalability of EWS. This flexibility ensures that these systems can effectively harness the immense and varied computational power provided by initiatives like the ECP and EuroHPC JU, ultimately leading to more robust and future-proof solutions. 

Currently, the only fully performance portable, scalable SWE solver for flood simulation available is SERGHEI-SWE \cite{SERGHEI}. SERGHEI leverages the Kokkos performance portability abstraction layer \cite{kokkos1,kokkos2} to enable GPU acceleration in combination with distributed programming with MPI. SERGHEI, by using Kokkos, is CUDA, OpenMP, HIP, SYCL, and Pthreads-ready. While SERGHEI has this advantage of portability, performance and computational efficiency remains a crucial part in a SWE solver dedicated to flash flood modeling. 

\subsection{Related Work}
\begin{figure}
        \centering
        \includegraphics[width=\linewidth]{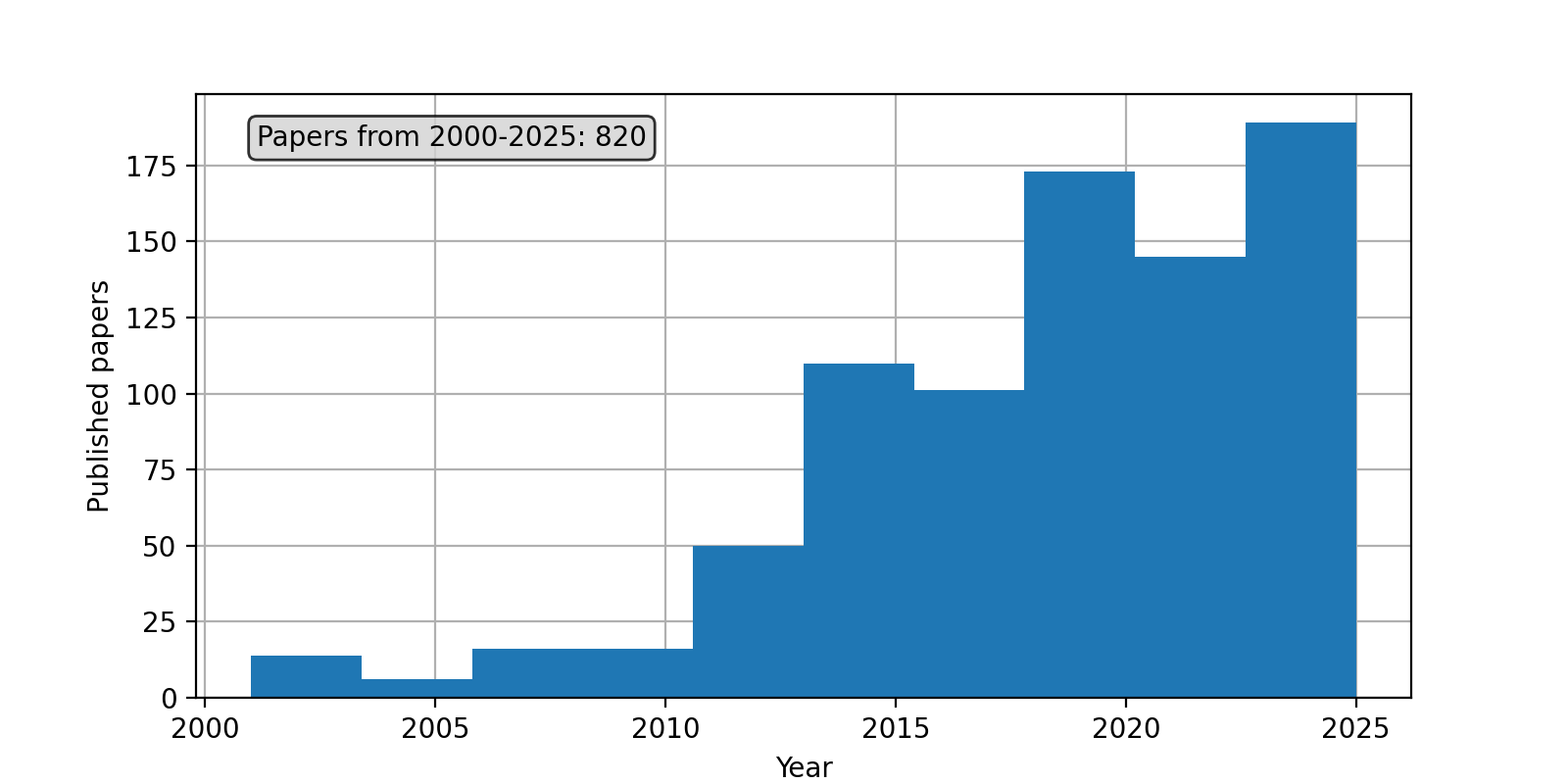}
        \caption{\centering Distribution of published papers (SCOPUS) from 2000 to 2025 containing performance portability in their title or abstract. Only computer science publications were considered.}
        \label{fig:perfport}
    \end{figure}

    Performance portability has been a topic of research since the 2000s, and it gained more interest among the computer science community since 2010, see Figure \ref{fig:perfport}. The introduction of better processors and the general-purpose GPU enabled the construction of different HPC systems with the end goal to create exascale systems \cite{exascale_project}. Performance portability libraries like Kokkos, RAJA, and OCCA have then been developed and studied extensively to evaluate their performance both in framework testing environments and large-scale scientific applications. 

    For example, in \cite{martin2025performance}, the HARVEY massively parallel multiphysics code undergoes a comparative performance analysis of the CUDA, SYCL, HIP, Kokkos, and OpenMP programming paradigms in leadership systems, Frontier, Aurora, Polaris, that contain devices from the major GPU vendors NVIDIA, AMD, and Intel. Their main contribution is the quantitative comparison of all the performance portability frameworks and key insights into the current state of HPC programming models. An interesting finding is that the SYCL programming paradigm showed itself as the most performance portable programming model on both CPU and GPU. A key takeaway from their work is that HPC developers are no longer limited neither by vendor-specific solutions nor by performance portable paradigms when implementing their HPC applications.

    Another recent effort was made in \cite{Bartolomeu2025}, by implementing an N-body problem in the following frameworks, OpenMP, pSTL, Kokkos, CUDA, HIP, SYCL, and OpenACC. The implementations were tested across four different types of GPU architectures and evaluated using two different performance portability metrics based on application efficiency normalized to the best performance on each platform. The metrics used are based on arithmetic means, reflecting recent advancements in the field as established in \cite{Marowka2021, Marowka2022, Marowka2025_1, Marowka2025_2} which question the initial harmonic mean approach \cite{pennycook_metric_2016,pennycook_revisiting_2021}. Key outlooks from this work refer to Kokkos being a better suited platform for complex memory access pattern GPU algorithms and OpenMP being a close to perfect portability layer according to their results.

    In the context of the SWE there have been efforts to design software to test different architectures and programming paradigms such as in \cite{Bttner2024}, in which the SWE were solved using discontinuous Galerkin methods using the SYCL programming model for portability. They evaluated the code on CPU, GPU and FPGA which is a strength of this paradigm. Their performance portability evaluation did not implement any of the metrics used currently, however they reported architectural and application performance for which these metrics can be calculated. By using roofline models they assure their implementation lies close to the performance peaks of CPUs and GPUs. FPGA execution is assured to surpass CPU and GPU performance when data fits in their generated caches. 

    Within this same context, in \cite{Elliott2018} the SWE are solved over a sphere using the radial basis function finite difference method as a simplified atmospheric dynamics code. They implemented the method using OpenMP and OpenACC combined with MPI to test both portability and scalability using Intel CPUs and NVIDIA GPUs. This work lacks tests on other GPU and CPU architectures and performance portability metric evaluations, yet they deemed their code performance portable as it achieves near theoretical peak performance on all platforms. 

    
\subsection{Contributions}

In this work we conduct a performance analysis to assess the SERGHEI-SWE solver \cite{SERGHEI}, under different GPU architectures focusing on the most popular vendors NVIDIA, AMD and Intel. 
The goal of the study is to
\begin{inparaenum}[(i)]
\item provide a performance evaluation of the SERGHEI-SWE solver in terms of scalability, portability and GPU kernel execution;
\item gain insights into the similarities and differences in performance and overall behaviour of the solver in different GPU architectures;
\item provide evidence and practical information for users on what to expect when migrating from one architecture or system to another, in terms of runtime and scalability.
\end{inparaenum}

We consider our efforts and results both important and novel, as they contribute to the growing body of knowledge on performance portability in scientific applications, and represent one of the first, if not the first, studies to evaluate the performance portability of a production-ready, highly scalable shallow water solver.

This work will be organized as follows:
\begin{itemize}
    \item Section \ref{sec:back} discusses the  background of this work.
    \item Section \ref{sec:method} explains the methodology used to evaluate the SERGHEI-SWE solver.
    \item Section \ref{sec:results} shows the experimental results obtained in this work.
    \item Section \ref{sec:final} refers to recommendations and future work for our study.
\end{itemize}

\section{Background}\label{sec:back}

\subsection{Numerical Method}
    The SERGHEI code solves for the Cartesian 2D SWE specifically, 
    \begin{align}
        \begin{split}
        \frac{\partial \mathbf{U}}{\partial t} + \frac{\partial \mathbf{F}}{\partial x} +\frac{\partial \mathbf{G}}{\partial y} = \mathbf{S}_\mathrm{r} + \mathbf{S}_\mathrm{b} + \mathbf{S}_\mathrm{f},&\\
        \mathbf{U} = 
        \begin{bmatrix}
            h \\ hu \\ hv
        \end{bmatrix}\,
        \mathbf{F} = 
         \begin{bmatrix}
            hu \\ hu^2+\frac{1}{2}gh^2 \\ huv    
        \end{bmatrix}\,
        \mathbf{G} = 
        \begin{bmatrix}
            hv \\ huv \\ hv^2 +\frac{1}{2}gh^2    
        \end{bmatrix},&\\
        \mathbf{S}_\mathrm{r} = 
        \begin{bmatrix}
            r_o - r_f \\ 0 \\ 0    
        \end{bmatrix}\,
        \mathbf{S}_\mathrm{b} = 
        \begin{bmatrix}
            0 \\ -gh \frac{\partial z}{\partial x} \\ -gh \frac{\partial z}{\partial y}
        \end{bmatrix}\,
        \mathbf{S}_\mathrm{f} = 
        \begin{bmatrix}
            0 \\ -\sigma_x \\ -\sigma_y    
        \end{bmatrix},&
        \end{split}
        \label{eq:SWE}
    \end{align}

    \noindent in which $h$ is the water height over the surface being simulated, $u$ and $v$ are the fluid velocity components in the $x$ and $y$ respectively, and $g$ is gravity. The source terms $\mathbf{S}_\mathrm{r}$, $\mathbf{S}_\mathrm{b}$, and $\mathbf{S}_\mathrm{f}$ take into account infiltration and exfiltration, the slope of the domain, and the friction of the terrain over the fluid being modeled. SERGHEI's numerical method starts with the integral over a control volume $\Omega$ of the equations,    
    \begin{equation}
        \frac{\partial}{\partial t} \int_{\Omega} \mathbf{U} d\Omega + \oint_{\partial\Omega} (\mathbf{E}\cdot \mathbf{n})dl =\int_{\Omega} (\mathbf{S}_{\text{r}}+\mathbf{S}_{\text{b}}+\mathbf{S}_{\text{f}}
        )d\Omega,
    \end{equation}
    \noindent in which $\mathbf{E}=(\mathbf{F},\mathbf{G})$ is the convective flux vector and $\mathbf{n}=(n_x,n_y)$ is the outward unit normal vector. The finite volume method equation is then derived by discretizing over the edges of the cell and considering a piecewise representation of the conserved variables,
    \begin{equation}
            \frac{\partial}{\partial t} \int_{\Omega_i} \mathbf{U} d\Omega + \sum_{k=1}^4 (\mathbf{E}\cdot \mathbf{n})_k l_k = \int_{\Omega_i} \mathbf{S}d\Omega+\int_{\Omega_i}\mathbf{S}_{\text{r}}d\Omega.
        \label{eq:FVM}
    \end{equation}
    \noindent Here $\mathbf{S}=\mathbf{S}_{\text{b}}+\mathbf{S}_{\text{f}}$, and the source term $\mathbf{S}_{\text{r}}$ is separate since SERGHEI uses a different numerical scheme for integration. The fluxes $\mathbf{E}\cdot \mathbf{n}$ can be found by applying the theory of approximate Riemann solvers \cite{Murillo2010, MURILLO20126861}. SERGHEI specifically implements an augmented solver such that the source terms provide a contribution to the fluxes at each interface, this ensures well-balancing in the solution.
    
    To find these fluxes, equation (\ref{eq:SWE}) can be projected onto $\mathbf{n}$ such that a one dimensional Riemann problem can be defined on the rotated set of coordinates $(\hat{x}, \hat{y})$ that lie on the normal and tangential directions of each edge. An approximate solution $\widehat{\mathbf{U}}(\hat{x}, t)$ can be defined and its integral over a defined control volume must be equal to the integral of the exact solution $\mathbf{U}(\hat{x}, t)$ over the same control volume \cite{Murillo2010}. Following this property, time linearization and projection of the source term onto the edge, equation (\ref{eq:FVM}) turns into 
    \begin{equation}
        \frac{\partial}{\partial t} \int_{\Omega_i} \mathbf{U} d\Omega + \sum_{k=1}^4 (\delta\mathbf{E}-\mathbf{T})_k\cdot\mathbf{n}_k l_k = \mathbf{0}
    \end{equation}
    \noindent where $\mathbf{S} = \mathbf{T}_k \cdot \mathbf{n}_k$ for an appropriate source term matrix $\mathbf{T}_k$, and the fact that $\sum_{k=1}^{4}\mathbf{n}_kl_k=0$ helps define $\delta \mathbf{E}=\mathbf{E}_j-\mathbf{E}_i$ in the sum. This problem can be solved approximately with, 
    \begin{align}
        \begin{split}
            & \frac{\partial}{\partial t} \int_{\Omega_i}\widehat{\mathbf{U}}d\Omega + \sum_{k=1}^4\widehat{\mathbf{J}}_{\mathbf{n},k}\delta\widehat{\mathbf{U}} l_k= \mathbf{0} \\
            &\widehat{\mathbf{U}}(\hat{x},0) = 
            \begin{cases}
                \mathbf{U}_i\quad\text{if $\hat{x} > 0$} \\
                \mathbf{U}_j\quad\text{if $\hat{x} < 0$,}
            \end{cases}
        \end{split}
        \label{eq:RPFVM}
    \end{align}
    
    \noindent in which an approximation of the flux $(\delta\mathbf{E}-\mathbf{T})_k\cdot\mathbf{n}_k = \widehat{\mathbf{J}}_{\mathbf{n},k}\delta\hat{\mathbf{U}}_k$ was applied. The Jacobian $\widehat{\mathbf{J}}_{\mathbf{n},k}$ can be approximated locally by using $\widetilde{\mathbf{J}}_k = \partial \mathbf{F(\hat{\mathbf{U}})}/\partial \hat{\mathbf{U}}$ evaluated at the Roe states. The analysis of this eigenstructure is useful to define an basis onto which the wave and source terms can be projected, resulting in,
    \begin{equation}
        \widetilde{\mathbf{J}}_k = \widetilde{\mathbf{P}}_k\widetilde{\mathbf{\Lambda}}_k\widetilde{\mathbf{P}}_k^{-1};\quad\delta\widehat{\mathbf{U}}=\widetilde{\mathbf{P}}_k\mathbf{A}_k;\quad \mathbf{T}\cdot\mathbf{n}_k=\widetilde{\mathbf{P}}_k\mathbf{B}_k,
    \end{equation}

    \noindent where $\mathbf{\Lambda}_k$ is the matrix of the eigenvalues of $\widetilde{\mathbf{J}}_k$ and $\widetilde{\mathbf{P}}_k=(\widetilde{\mathbf{e}}_1,\widetilde{\mathbf{e}}_2,\widetilde{\mathbf{e}}_3)$ the matrix made up of its eigenvectors. Some algebra leads to define the fluxes as, 
    
    \begin{equation}
        \widetilde{\mathbf{J}}_k\delta\widehat{\mathbf{U}}_k- \mathbf{T}_k\cdot\mathbf{n}_k = \sum_{m=1}^{3}\left[(\tilde{\lambda}\alpha-\beta)\tilde{\mathbf{e}}\right]_{m,k}
        \label{eq:fluxes_lambdas}
    \end{equation}
    
    \noindent where $\tilde{\lambda}_m$ are the elements of $\mathbf{\Lambda}_k$ and  $\alpha_m, \beta_m$ the elements of $\mathbf{A}_k$ and $\mathbf{B}_k$ respectively. The approximate problem is integrated in time using a forward Euler method, substituting equation (\ref{eq:fluxes_lambdas}) into (\ref{eq:RPFVM}), assuming a cell averaged approximate solution, and performing upwinding in the discretization, the update rule for the conserved variables $\mathbf{U}$ is,
    \begin{equation}
        \mathbf{U}_i^{n+1} = \mathbf{U}_i^{n} - \frac{\Delta t}{\Delta x}\sum_{k=1}^{4}\sum_{m=1}^{3}\frac{\tilde{\lambda}^{-}}{\tilde{\lambda}}\left[(\tilde{\lambda}\alpha-\beta)\tilde{\mathbf{e}}\right]_{m,k}^{n} + [\mathbf{S}_\text{r}]_i^n \Delta t
        \label{eq:stateupdate}
    \end{equation}
    \noindent with a cell evaluated Courant-Friedrichs-Lewy (CFL) condition, 
    \begin{equation}
        \Delta t = \text{CFL} \min_{i}{\left\{ \frac{\Delta x}{|u_i|+\sqrt{gh_i},|v_i|+\sqrt{gh_i}} \right\}}
        \label{eq:timestep}
    \end{equation}

\subsection{Implementation}

    SERGHEI implements a distributed memory approach using MPI to divide the Cartesian computational domain across a Cartesian grid of processes. At initialization, the $(N_x,N_y)$ domain grid is mapped onto a $(P_x,P_y)$ process grid such that each process handles a local subgrid of size $(n_x+2, n_y+2)$, including halo cells for communication. The process $(P_i,P_j)$ in the grid is defined by $(P_i = R \% P_t, P_j = R / P_t)$ where $R$ is the process rank number and $P_t=P_xP_y$ the total number of processes.

    The fields in the local subgrid of size $(n_x+2, n_y+2)$ are represented in SERGHEI as 1D arrays of size $N_{\text{arr}}=(n_x+2)(n_y+2)$, with indexing $k=j(n_y+2)+i$. 
    To fully exploit modern HPC resources, SERGHEI leverages the Kokkos library for shared memory and GPU parallelism using simple \texttt{Kokkos::parallel\_x} concepts. These field arrays are implemented using the \texttt{Kokkos::View} data structure, ensuring efficient data placement close to where computations occur. By adopting a unified memory paradigm across all device backends, Kokkos abstracts explicit host-to-device memory transfers, allowing execution across heterogeneous architectures.
    
    To ensure portability across GPU architectures, SERGHEI explicitly uses the \texttt{Kokkos::SharedSpace} and \texttt{Kokkos::DefaultExecutionSpace} concepts as these translate to the unified memory spaces and execution environments defined by the Kokkos configuration. For instance, on NVIDIA devices, they will map to \texttt{Kokkos::CudaUVMSpace} and \texttt{Kokkos::Cuda} respectively.

    With this data partition and representation, SERGHEI follows a structured execution flow for parallel computation of the numerical solution. The simulation begins with an initialization phase, where input data is parsed from model parameter files and loaded in parallel from NetCDF files using the PNetCDF library. These files define key simulation parameters, set boundary conditions, and initialize solution fields. Depending on the input type, SERGHEI either restores a previous state from a checkpoint or constructs initial conditions from scratch.  
    
    Once initialized, the simulation advances through a structured loop that iteratively computes the numerical solution. Each iteration follows a defined sequence: computing the time step using equation (\ref{eq:timestep}), approximating fluxes with equation (\ref{eq:fluxes_lambdas}), updating the state to $\mathbf{U}_i^{n+1}$, applying boundary conditions, exchanging domain halos for wet-dry cell corrections, ensuring continuity by exchanging the $\mathbf{U}$ state across domains, accounting for additional mass and momentum introduced by the boundary conditions, and performing synchronous parallel I/O  at user-specified times. The simulation finalizes by writing log files and finalizing the Kokkos and MPI runtimes.

    Regarding the simulation loop, the most computationally intensive kernels are specifically the ones pertaining to the calculation of the edge fluxes, which are implemented for the $x$ and $y$ coordinates separately to avoid repeated calculations, the time step computation kernel, and the state update kernel. 

\begin{figure}
        \centering
        \includegraphics[width=0.7\linewidth]{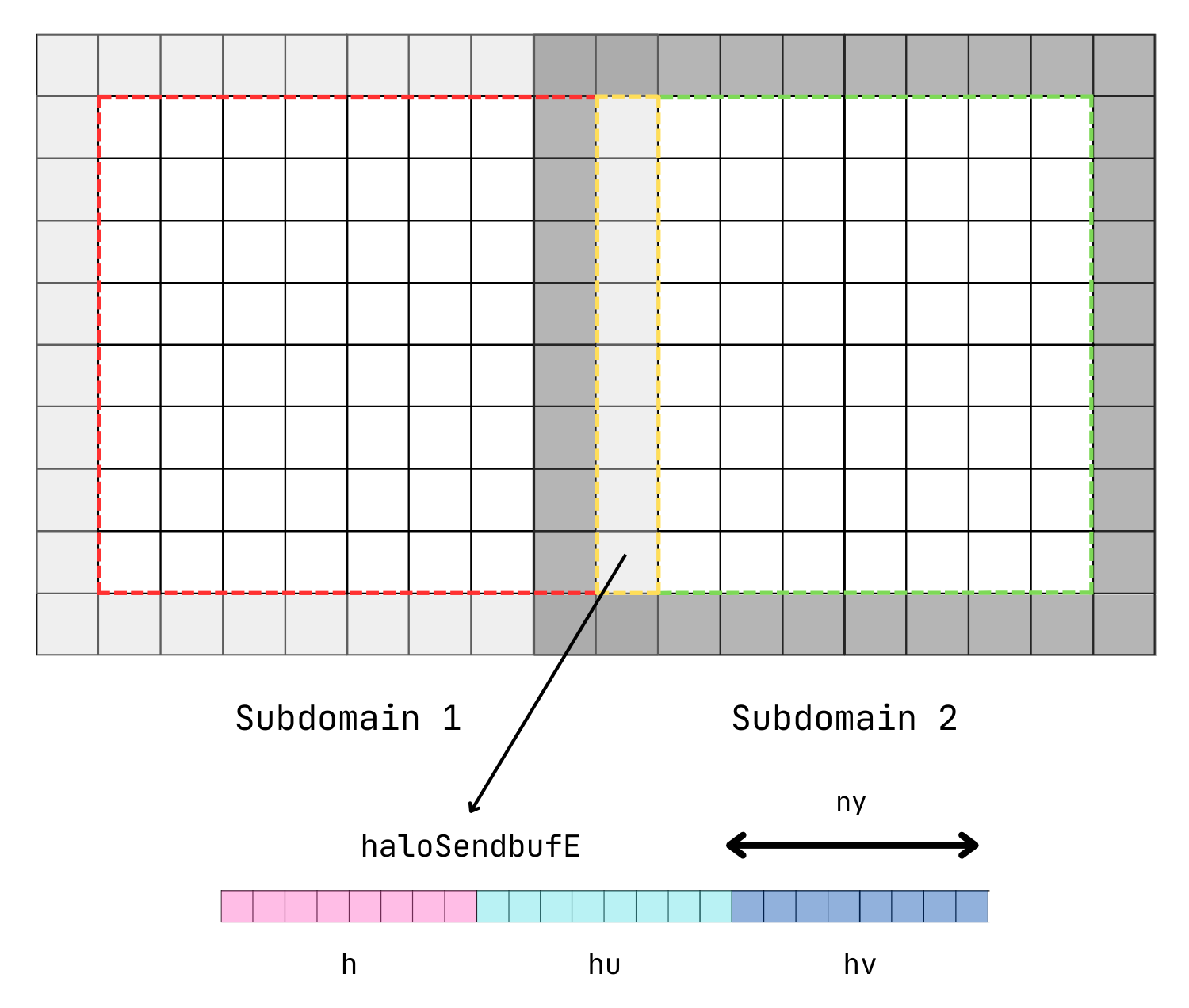}
        \caption{\centering Halo exchange packaged buffers for left-right communication between subdomains.}
        \label{fig:comms}
    \end{figure}

    Parallel communication is implemented using asynchronous send/receive operations of a packed halo buffer which groups the state variables. The order in which they are grouped is water depth $(h)$, horizontal unit discharge $(hu)$ and vertical unit discharge $(hv)$. Figure \ref{fig:comms} shows a diagram of the communication buffers for left-right communication. 
    
    The kernels offloaded to the GPU follow simple Kokkos parallel dispatch patterns such \texttt{Kokkos::parallel\_for} and \texttt{Kokkos::parallel\_reduce} with 1D loops specified by either the total number of cells in the subdomain or the number of valid cells in the subdomain (not including the halo cells).

\section{Methodology}\label{sec:method}

This section details the methodology used to assess the computational performance of the SERGHEI code. Our analysis focuses on a key test: a circular dam break hydrological simulation. This case is perfect to characterize the performance of code in ideal conditions. 



\subsection{Test Case: circular dam break} 

    The circular dam break is a generalisation from the one dimensional dam break to two dimensions by implementing a full rotation along a $z$ axis. This case is particularly beneficial as it has regular, uniform dynamics such that load balancing is not an issue when performing scaling tests. Our circular dam break case is then set up as a square domain of side $N_\text{side}\Delta x$ where $N_\text{side}$ is the number of cells in that specific side and $\Delta x=0.5$. The initial conditions then can be set as,
    \begin{equation}
        h(x,y,t=0) = 
        \begin{cases}
            4&\text{if $r=\sqrt{x^2+y^2} \leq N_\text{side}\Delta x/5$} \\ 
            1&\text{otherwise.} \\ 
        \end{cases}
    \end{equation}

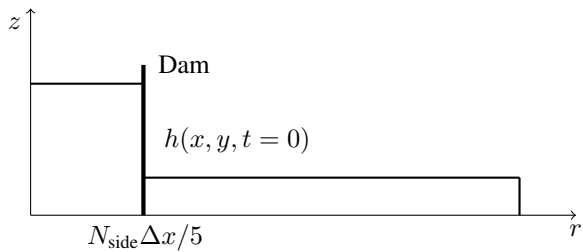
\begin{figure}
    \centering
    \begin{tikzpicture}
            
        \draw[ultra thick] (1.5,0) -- (1.5,2.0);
        \draw[thick] (0,1.75) -- (1.5,1.75);
        \draw[thick] (1.5,0.5) -- (6.5,0.5);
        \draw[thick] (6.5,0) -- (6.5,0.5);
        \node[text width=2cm] at (1.75,-0.3) {$N_\text{side}\Delta x/5$};
    
    
        \draw[->] (0,0) -- (7.25,0) node[anchor=north] {$r$};
        \draw[->] (0,-0) -- (0,2.75) node[anchor=north east] {$z$};
        \node at (2.75, 1) {$h(x,y,t=0)$};
        \node[text width=2cm] at (2.7, 2.0) {Dam};
    
    \end{tikzpicture}
    \caption{\centering Diagram of the initial conditions of the implemented dam break test case in the $rz$ plane.}
    \label{fig:dambreakdiagram}
\end{figure}

    \noindent Figure \ref{fig:dambreakdiagram} shows the initial conditions on the $rz$ plane. The boundary conditions are defined to be reflective at the boundaries. Our implementation is dependent on the number of cells $N_\text{side}$, the number of processes in the $x$ and $y$ dimensions of the cartesian topology ($P_x$, $P_y$) the simulation length $t_{\text{sim}}$ and the output frequency $t_{\text{I/O}}$. We vary these parameters according to the different studies we carry out.

    \vspace{0.2in}

\subsection{Systems and Configuration}
    Our performance analysis relies on four HPC supercomputing systems, each with a distinct accelerator architecture. This diversity enables the evaluation and execution of code with the MPI+(CUDA, HIP, SYCL) programming paradigms through the Kokkos portability layer. Table \ref{tab:compilers} and \ref{tab:hardware} list the compilers and systems used in our testing and data acquisition phase as well as the software stack employed in the experiments. This includes the versions of MPI libraries, GPU programming environments (CUDA, HIP, SYCL), compiler toolchains, and additional dependencies required for Kokkos-based execution.

    \begin{table}
        \centering
        \caption{Compiler versions used for each code variant.}
        \begin{tabular}{c|c}
            \hline
            System & Compiler   \\
            \hline
            AURORA & Intel(R) oneAPI DPC++/C++ Compiler  \\ 
            & 2025.0.0 (2025.x.0.20240629) \\ 
            FRONTIER & AMD clang version 17.0.0 (roc-6.0.0 23483  \\ 
            & 7208e8d15fbf218deb74483ea8c549c67ca4985e)\\
            JUWELS BOOSTER & g++ (GCC) 12.3.0 \\ 
            JEDI & g++ NVHPC/24.9-CUDA-12 \\ 
            \hline
        \end{tabular}
        \label{tab:compilers}
    \end{table}

    \begin{table*}
        \centering
        \caption{HPC systems used to test the SERGHEI SWE solver.}
        \begin{tabular}{cccccccc}
            \hline
            System & Centre & CPU & GPU  & Vendor & On-Node Accelerators & Nodes                               \\
            \hline
            AURORA & ANL & Intel Sapphire Rapids & Max 1550 Series & Intel & $6\times(\text{2 Tiles})$ & 10624      \\
            FRONTIER & OLCF & AMD Optimized 3rd Gen EPYC & Instinct MI250X & AMD & $4\times(\text{2 GCD})$ & 9408   \\
            JUWELS BOOSTER & JSC & AMD EPYC 7402 & Ampere A100 & NVIDIA & 4 & 936 \\
            JEDI & JSC & NVIDIA Grace Arm Neoverse-V2 & Hopper H100 &  NVIDIA & 4 & 48 \\
            \hline
        \end{tabular}
        \label{tab:hardware}
    \end{table*}

    \begin{table*}
        \centering
        \caption{HPC software stack used to compile and test the SERGHEI SWE solver. All systems implemented Kokkos 4.5.99.}
        \begin{tabular}{cccccc}
            \hline
            System & MPI & Kokkos backend & PNetCDF  & Linux Version \\
            \hline
            AURORA          & MPICH/4.3.0rc3        & OneAPI-2025.x.0.20240629 & 1.12.3 & SUSE-5.14.21-150400.24.55-default  \\
            FRONTIER        & CRAY-MPICH/8.1.29     & ROCM-6.0.0               & 1.12.3 & SUSE-6.4.0-150600.23.17\_14.0.63-cray\_shasta\_c   \\
            JUWELS BOOSTER  & ParastationMPI/5.9.2-1& CUDA-12.2.91             & 1.12.3 & Rocky-5.14.0-503.23.1.el9\_5.x86\_64   \\
            JEDI            & NVHPC-OpenMPI/4.1.5   & CUDA-12.6.20             & 1.12.3 & Red Hat-5.14.0-503.26.1.el9\_5.aarch64+64k  \\
            \hline
        \end{tabular}
        \label{tab:software}
    \end{table*}
   
\subsection{Strong and weak scaling}

    We conducted strong scaling tests on each platform using a circular dam break scenario with $N_{\text{side}} = 36000$ cells, totaling $N_{\text{tot}} = 1.296 \times 10^9$ cells. The code was instrumented with the \texttt{Kokkos::timer} tool to measure execution times and \texttt{Kokkos::fence} to ensure correct GPU kernel measurements. 
    
    Only the simulation loop time was measured, as it is the computationally intensive part of the code (i.e., excluding initialization and I/O). For data analysis, we considered the number of physical GPUs per node rather than those detected by the resource schedulers, since some architectures feature multiple tiles or graphics compute dies (GCDs) per physical GPU. To reduce statistical variation, each execution was repeated five times. Parallel speedup was calculated as the ratio of the simulation time on a single node using four physical GPUs to the simulation time at larger scales, $s=t_\text{node}/t$ . In the case of Intel GPUs we add several points at multiples of four GPUs instead of the six GPUs featured by an Aurora node to compare data points across all architectures. 

    Regarding weak scaling tests, we kept the number of cells in the circular dam break constant per physical GPU, specifically $N_\text{weak}=1.28\times10^8$. The problem was scaled up to 1024 nodes where resources allowed. The cases where the physical GPU is divided into tiles or GCDs in the case of the Aurora or Frontier systems, this number is divided by the number of subchips in the GPU. Following this, we chose $N_\text{weak}=6.4\times10^7$ for each subchip. Given that we used the dam break case with $N_\text{weak} = N_\text{side}^2$ it is evident that $N_\text{side}$ is not necessarily an integer, in these cases we considered $N_\text{weak} = (N_\text{side}+1)^2$. After measurement, the efficiency was quantified with respect to the ratio of simulation time at large scales to the simulation time on a single node using four physical GPUs, $e=t_\text{node}/t$. For the architectures that feature 6 physical GPUs per node this might imply that some efficiencies might surpass the ideal efficiency.


\subsection{Roofline analysis}
       
    
    We focused on profiling key offloaded kernels, particularly those responsible for the most computationally intensive operations in our numerical method. Specifically, we analyzed the \texttt{computeDt} kernel, which is associated with equation (\ref{eq:timestep}) for timestep computation. We also measured \texttt{computeDeltaFluxXRoe} and \texttt{computeDeltaFluxYRoe}, which correspond to equation (\ref{eq:fluxes_lambdas}) for flux computations. Additionally, we profiled \texttt{computeNewState}, which updates the solution state following equation (\ref{eq:stateupdate}), and \texttt{computeTimeStepReduction}, which applies a timestep correction for wet-dry interfaces.
    
    To perform this analysis, we employed NVIDIA Nsight Compute, AMD ROCm Profiler, and Intel Advisor. Following the methodology established in \cite{yang2020hierarchicalrooflineanalysiscollect}, we constructed roofline graphs for each GPU architecture tested. We then normalized the results for each roofline by peak floating operations and by an arithmetic intensity threshold to compare all architectures and their behavior in a single graph. The normalizations have this form,
    \begin{equation}
        p_\text{norm} = \frac{p_\text{achieved}}{p_\text{peak}};\quad\quad a_\text{norm}=\frac{a_\text{achieved}}{a_\text{thresh}};\quad\quad a_\text{thresh}=\frac{p_\text{peak}}{b_\text{peak}}    
        \label{eq:rooflinenorm}
    \end{equation}
    \noindent where $p$ is processing speed measured in FLOP/s, $a$ is arithmetic intensity measured in FLOP/Byte and $b$ is memory bandwidth measured in Byte/s. The peak values were calculated empirically on NVIDIA and AMD architectures using the Empirical Roofline Tool (ERT) \cite{ERT} and using Intel Advisor on Intel architectures. ERT is a toolkit that launches simple $d=ab+c$ operations to quantify real values for peak processing speed and peak bandwidth in specific hardware.

The circular dam break application was evaluated using two different configurations based on problem size: one with a fixed problem size ($N_x=15000 $) and another with the size that yielded the best performance. To determine the optimal configuration, we tested a range of problem sizes, varying $N_x$ from 4000 to 20000.

\subsection{Performance Portability Metrics}
    To quantify performance portability across architectures, we employed several Performance Portability ($\ppmetric$) metrics as proposed in \cite{Marowka2021, Marowka2022, pennycook_metric_2016, pennycook_revisiting_2021}. The metrics we calculated take the following forms:
    \begin{align}
        \begin{split}
            \ppmetric_1(\alpha, \varphi, \mathcal{H}) &= 
            \begin{cases}
                \displaystyle\frac{|\mathcal{H}|}{\sum_{i \in \mathcal{H}} \displaystyle\frac{1}{r_i(\alpha,\varphi)}} \quad \text{if}\quad \forall i \in \mathcal{H}, r_i\neq 0 \\
                0 \quad \text{otherwise},
            \end{cases} \\
            \ppmetric_2(\alpha,\varphi, \mathcal{H}) &= 
            \begin{cases}
                \displaystyle\frac{\sum_{i \in \mathcal{H}} r_i(\alpha,\varphi)}{\mathcal{|H|}} \quad \text{if}\quad |\mathcal{H}| \neq 0 \\
                0 \quad \text{otherwise},
            \end{cases} \\
        \end{split}
    \end{align}
    where $r_i(\alpha, \varphi)$ corresponds to the relative performance on platform $i\in\mathcal{H}$ of a specific application $\alpha$ for a specific problem $\varphi$. It is calculated as,
    \begin{equation}
        r_i = \frac{p_\text{achieved}}{\min{\{p_\text{peak}, b_\text{peak}a_\text{achieved}}\}}
    \end{equation}
    to account for both compute bound and memory bound kernels.

    As we do not have low-level, highly optimized versions of our application for all architectures, we solely consider architectural efficiency as recommended in \cite{Marowka2022}. We computed these metrics to obtain a scalar value in the range $[0,1]$, where a value of 1 indicated full portability with optimal efficiency across all platforms. We computed these metrics using the data collected for the roofline analysis both for uniform problem sizes and best performing problem sizes.
\section{Results and Discussion}\label{sec:results}
\begin{figure*}
    \begin{subfigure}[t]{0.4875\textwidth}
        \centering
        \includegraphics[width=\linewidth]{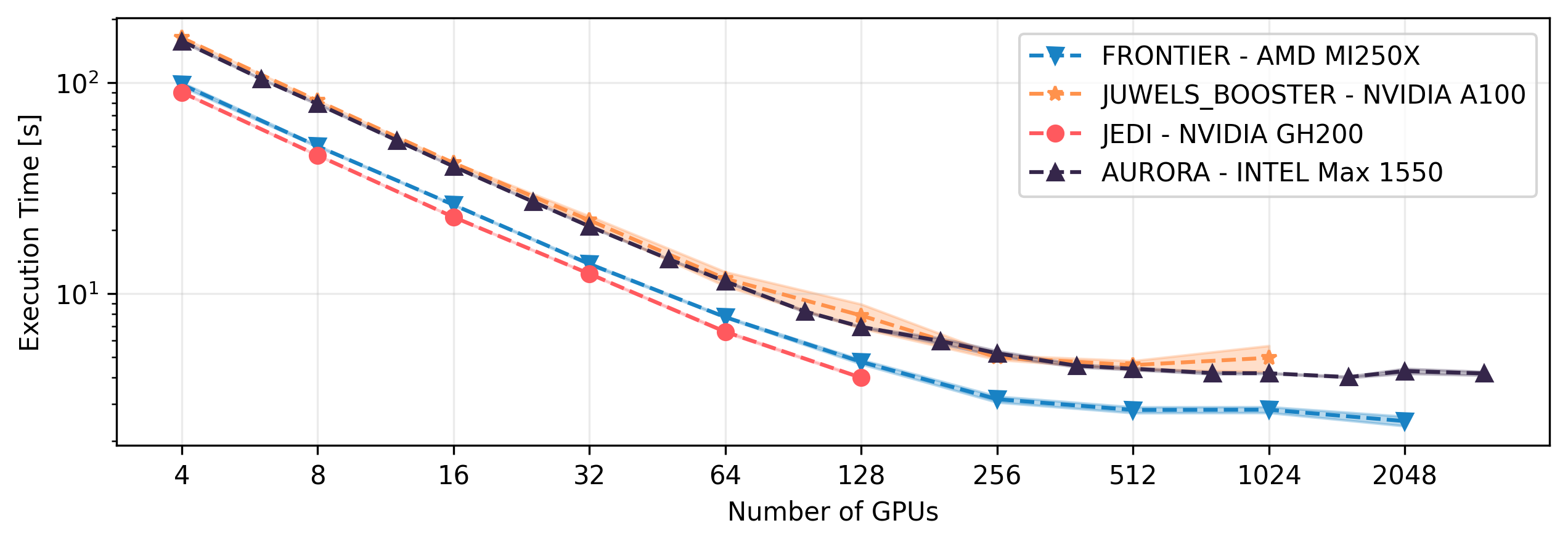}
        \caption{\centering Strong scaling execution time results for all HPC systems for the dam break case with $N_\text{side}=36000$.}
        \label{fig:exectime}
    \end{subfigure}\hfill
     \begin{subfigure}[t]{0.49\textwidth}
        \centering
        \includegraphics[width=\linewidth]{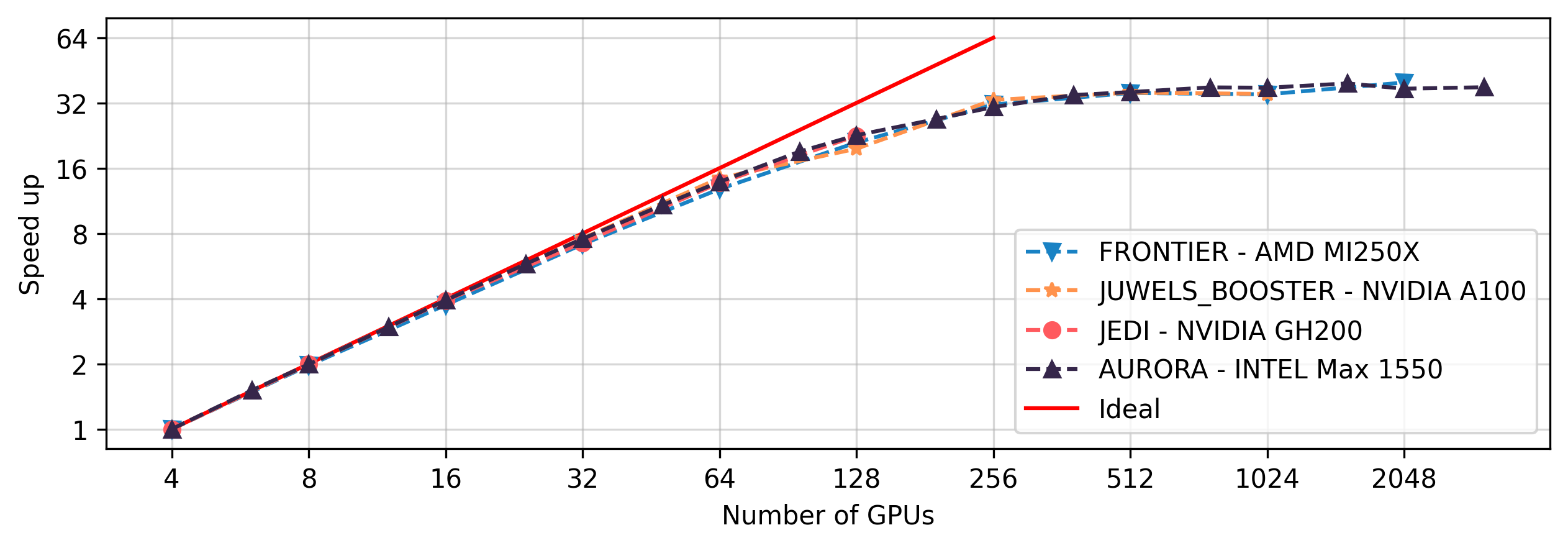}
        \caption{\centering Speedup results for all HPC systems for the dam break case with $N_\text{side}=36000$.}
        \label{fig:strongscaling}
    \end{subfigure}\hfill
    \begin{subfigure}[t]{0.49\textwidth}
        \centering
        \includegraphics[width=\linewidth]{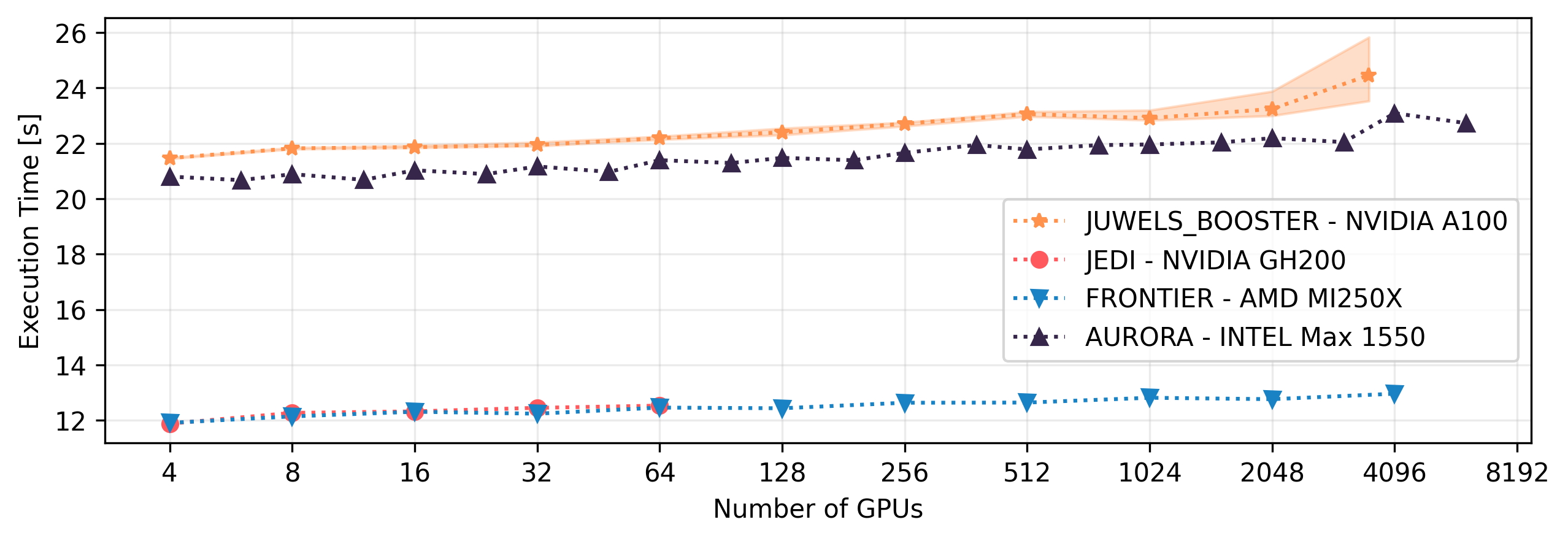}
        \caption{\centering Weak scaling execution time results for all HPC systems along with total amount of computational cells.}
        \label{subfig:weakscalingtime}
    \end{subfigure} \hfill
    \begin{subfigure}[t]{0.49\textwidth}
        \centering
        \includegraphics[width=\linewidth]{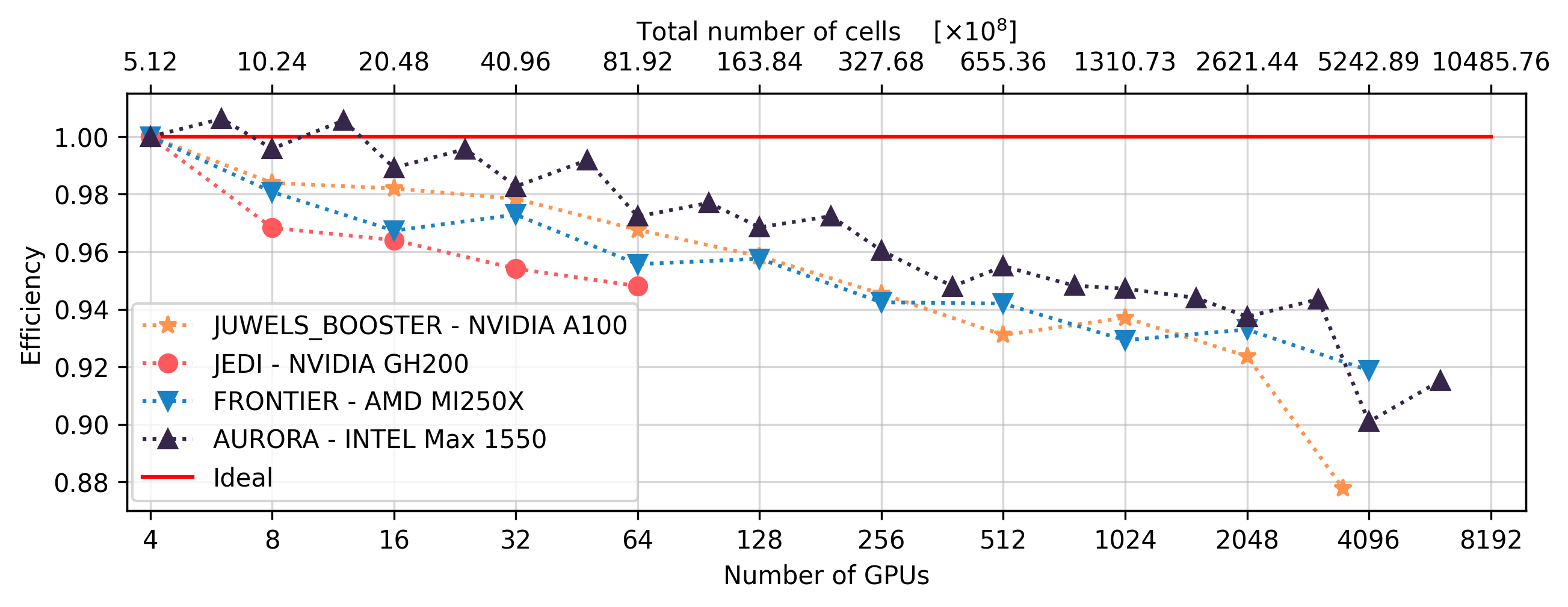}
        \caption{\centering Weak scaling efficiency results for all HPC systems along with total amount of computational cells..}
        \label{subfig:weakscaling}
    \end{subfigure} \hfill
    \caption{\centering Strong and weak scaling results for all the tested architectures. Weak scaling results were collected up to 1024 computing nodes for the dam break case (in systems that supported this amount of nodes).} 
    \label{fig:scalingresults}
\end{figure*}

\subsection{Strong and Weak Scaling Analysis}
    Strong scaling results demonstrate efficient parallel performance across all tested architectures up to approximately 512 to 1024 GPUs, as shown in Figure \ref{fig:exectime}. Among the platforms, JEDI’s NVIDIA GH200 system achieves the shortest execution times across all configurations, closely followed by Frontier's AMD MI250X. Interestingly, Aurora’s Intel Max 1550 and JUWELS Booster’s NVIDIA A100 exhibit comparable execution times, suggesting similar performance characteristics despite differences in underlying hardware and GPU architectures. These trends highlight the impact of platform-specific features, such as memory bandwidth, interconnect topology, and GPU architecture, on the scalability of large-scale simulations.
    
    Speedup results, shown in Figure \ref{fig:strongscaling}, confirm that the SERGHEI-SWE code scales as expected on all platforms plateauing at about 32 times single node time. These results exhibit near-linear scaling behavior up to 32-64 GPUs. As the number of GPUs increases, deviations from ideal speedup emerge, which can be attributed to increasing communication overheads and reduced per-GPU workload. Nonetheless, all platforms maintain a good degree of parallel efficiency, particularly in the low-to-mid GPU range.

    In practice, simulations with this solver generally do not exceed $N_{\text{tot}} = 1.296 \times 10^9$ cells. The circular dam break case serves as a well-defined analytical scenario, offering insight into the solver's behavior under ideal conditions. It provides a balanced framework for understanding how performance scales with increasing resources, establishing an upper bound for performance in the absence of load balancing issues.

    On the other hand, our weak scaling analysis allowed us to test the limits of the SERGHEI-SWE code, examining its behavior for extremely large problem sizes. In this regard, Figure \ref{subfig:weakscaling} presents the weak scaling efficiency for the circular dam break case across the four systems. All platforms maintain relatively high weak scaling efficiency up to 64 GPUs, with most above 95\%. However, as the GPU count increases beyond this point, we observe a gradual decline in efficiency, more pronounced for the JUWELS BOOSTER system. This behavior may be explained by increased communication overhead and topology related issues as we reach the maximum number of nodes in the JUWELS BOOSTER system.
    
    AURORA (Intel Max 1550) consistently delivers the highest efficiency across nearly the entire scaling range, particularly at large GPU counts (1024+), which highlights the strength of its interconnect architecture in sustaining load balance and minimizing communication bottlenecks. In contrast, JUWELS BOOSTER shows the steepest efficiency drop at higher scales, with performance falling below 90\% at 4096 GPUs.
    
    Even though the theoretical peak performance and memory bandwidth of JEDI's GH200 should favor weak scaling, its efficiency begins to trail off slightly earlier than AURORA and FRONTIER. This could reflect early-stage tuning, GPU underutilization, or network saturation issues (the reader should keep in mind that JEDI is the development instrument for the upcoming JUPITER system, and not yet the production machine).
    
    Overall, the results indicate that while all architectures demonstrate good weak scaling up to medium-large scales, maintaining high efficiency at extreme scales (2048+ GPUs) requires advanced interconnect performance and careful optimization of memory and communication patterns. These insights are valuable for informing future deployment strategies, targeting performance tuning efforts and optimization.

\subsection{Roofline Analysis}

    \begin{table}
        \centering
        \caption{\centering Empirical peak performance results obtained with ERT and Intel Advisor for a single GPU/GCD/Tile.}
        \scalebox{0.95}{\begin{tabular}{ccc}
            \hline
            \multirow{2}{*}{System}          & Theoretical Peak & Empirical Peak \\
            & \multicolumn{2}{c}{($p_\text{peak}$ [GFLOP/s] |$b_\text{peak}$ GB/s)} \\ 
            \hline
            AURORA          & (17000.00 | 3276.80) \cite{alcf_aurora_user_guide}    & (11400.00 | 1203.43) \\
            FRONTIER        & (23900.00 | 1600.00) \cite{ornlFrontierUser}          & (20105.84 | 1229.92) \\
            JEDI            & (34000.00 | 4000.00) \cite{nvidia_h100_datasheet}     & (31209.00 | 3032.59) \\
            JUWELS BOOSTER  & ( 9700.00 | 2039.00) \cite{nvidia_a100_datasheet}     & ( 9494.71 | 1258.40) \\
            \hline
        \end{tabular}}
        \label{tab:PeakPerf}
    \end{table}

Normalized roofline graphs were constructed as means of qualitative comparison of performance behavior across all architectures. Table \ref{tab:PeakPerf} summarizes the empirical peak values used to normalize both axes in the graphs.

\begin{figure}
    \begin{subfigure}{0.475\textwidth}
        \centering
        \includegraphics[width=\textwidth]{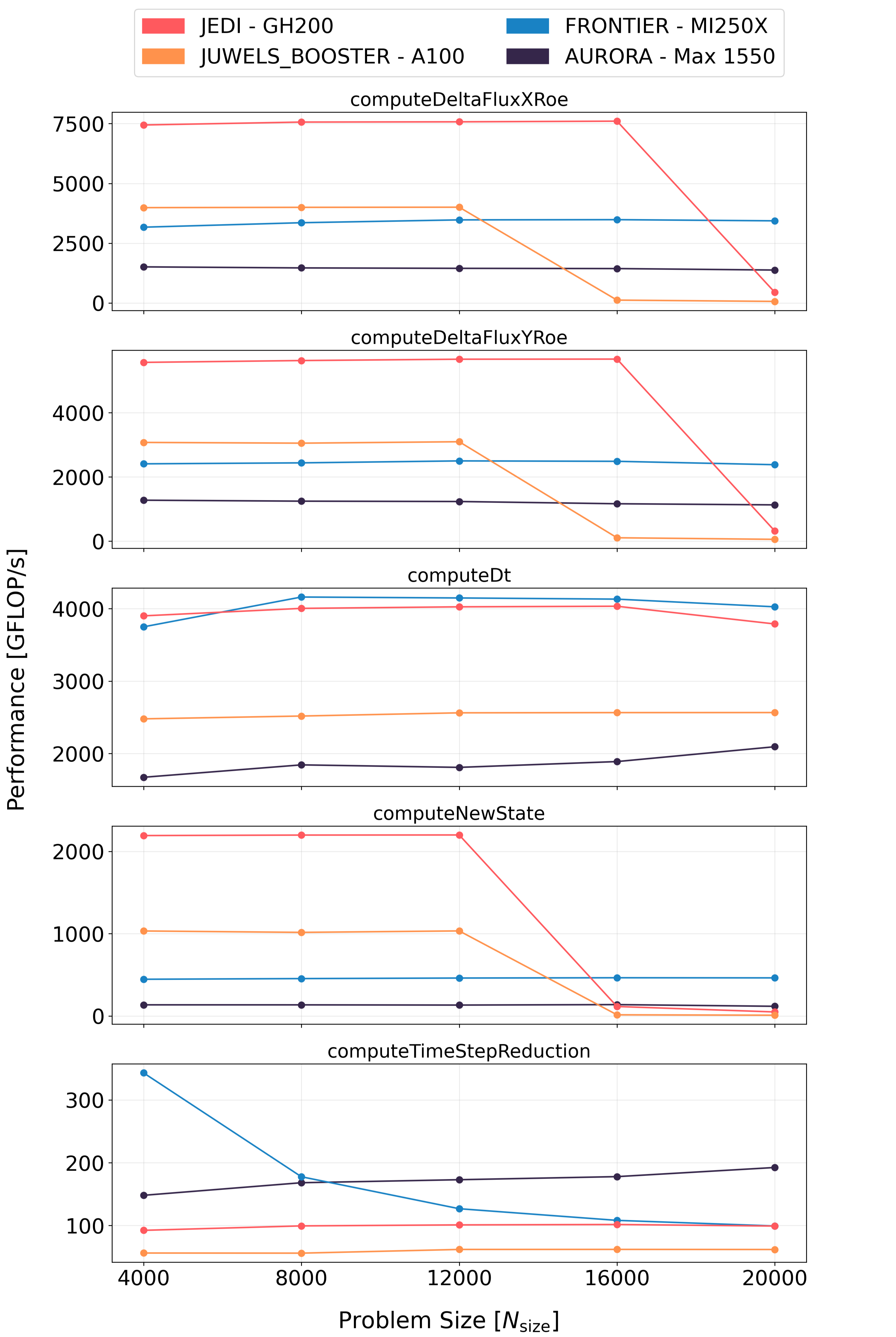}
        \caption{\centering Variation of performance with problem size for the analyzed GPU kernels. We hypothesize that drops may be caused by uncoalesced accesses in NVIDIA architectures.}
        \label{subfig:nxvariation}    
    \end{subfigure}\hfill
    \begin{subfigure}{0.475\textwidth}
        \centering
        \includegraphics[width=\textwidth]{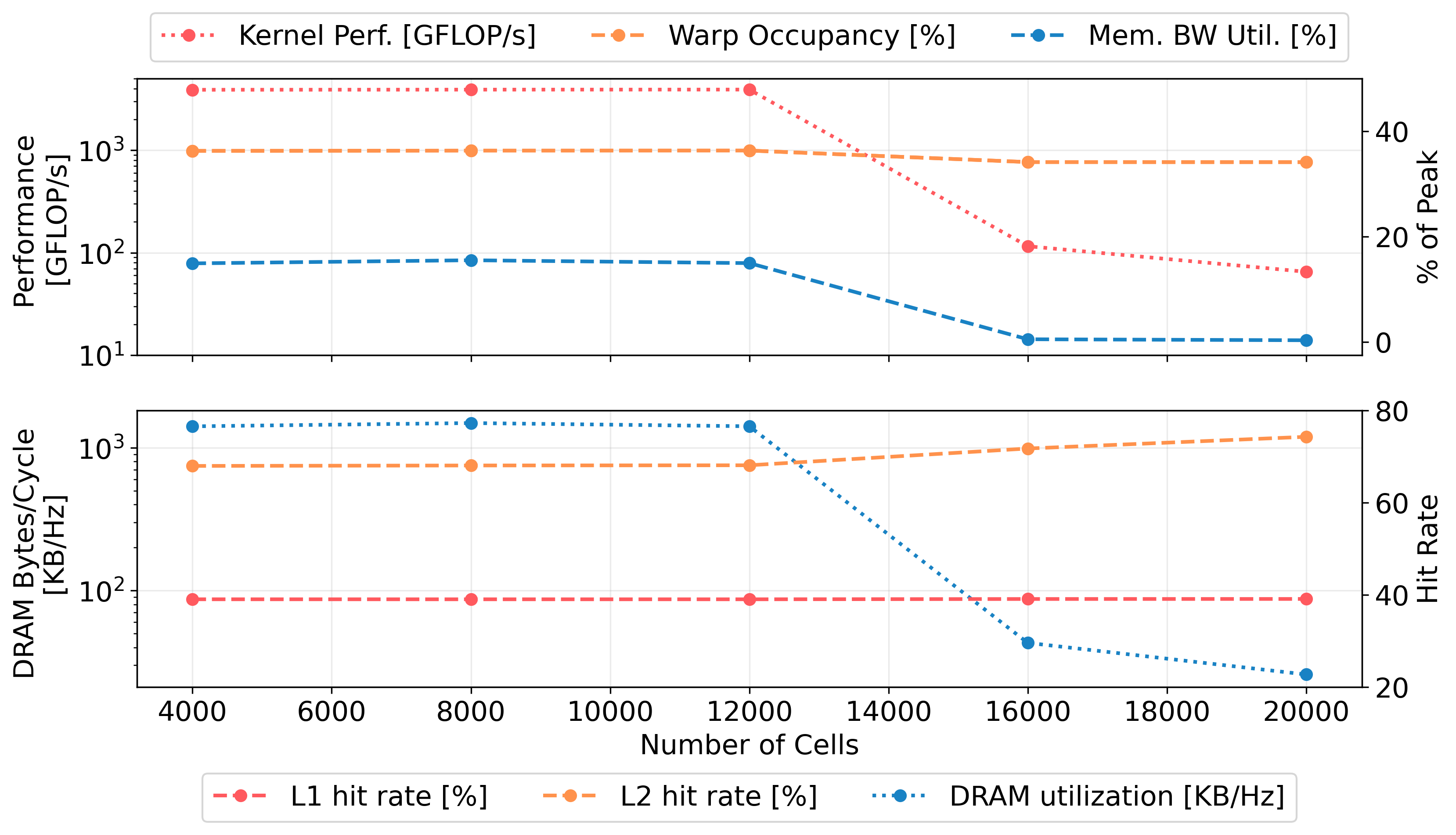}
        \caption{\centering Variation of profiled variables with problem size for the \texttt{computeDeltaFluxXRoe} Kernel on the JUWELS Booster platform. We hypothesize that drops may be caused by uncoalesced accesses in NVIDIA architectures.}
        \label{subfig:profiling}    
    \end{subfigure}\hfill
\end{figure}

Figure \ref{subfig:nxvariation} illustrates the performance variation of all profiled kernels on a single GPU. The \texttt{computeDeltaFluxXRoe} and \texttt{computeDeltaFluxYRoe} are the most expensive kernels in SERGHEI-SWE, typically demanding around 40\% of the simulation time, and thus we give them higher attention. As expected, performance increases with the number of cells across all architectures. However, performance plateaus around $10^7$ to $10^8$ cells. At larger problem sizes, the NVIDIA architectures exhibit an unusual sharp decline in performance.

Figure~\ref{subfig:profiling} shines a light on this specific issue. In a dedicated profiling run, we measured the \texttt{computeDeltaFluxRoe}'s kernel performance and other metrics as a function of problem size to discard certain theories that may have explained the performance drop. Despite L1 and L2 cache hit rates remaining relatively stable (approximately 40\% and 68–73\%, respectively), FLOP/s collapses at larger problem sizes, indicating that cache inefficiency is not the primary bottleneck. Warp occupancy decreases only slightly from 39\% to approximately 38\%, confirming that sufficient warps remain active and that SM starvation is not the dominant factor. In contrast, effective DRAM information movement (Bytes/cycle) drops sharply to near zero, correlating closely with the observed FLOP/s degradation. These results suggest that the kernel is primarily limited by memory latency, rather than raw bandwidth, a limitation likely caused by coalescing inefficiencies arising from our flattened 1D array indexing. This  behavior is likely to be propagating to all kernels since the code is based on this indexing scheme. It is important to note that this particular behavior does not happen on Intel and AMD architectures likely due to their different memory hierarchy and their ways of handling and optimizing low occupancy for kernels. 

We can also see an abnormal behavior in the performance of the  \texttt{computeTimeStepreduction} kernel, specifically for the FRONTIER system. Performance decreases with problem size. We hypothesize that the observed exponential performance decrease in the Frontier system for the \texttt{computeTimeStepReduction} kernel is attributed to a combination of architectural and algorithmic factors: (i) extremely low arithmetic intensity ($\sim 6$ FLOPS per thread) leading to memory-bound behavior, (ii) low occupancy due to the lightweight nature of the kernel relative to the 64-thread wavefront width, (iii) non-coalesced 1D memory accesses across large arrays, which penalizes memory throughput on GPU architectures, and (iv) diverging warps which is caused by a conditional statement within the kernel \cite{amd2024hip,amd2024hip_hw}. These factors compound as the problem size grows, explaining the kernel's performance degradation.

\begin{figure*}
    \centering
    \begin{subfigure}[t]{0.475\textwidth}
        \centering
        \includegraphics[width=0.975\linewidth]{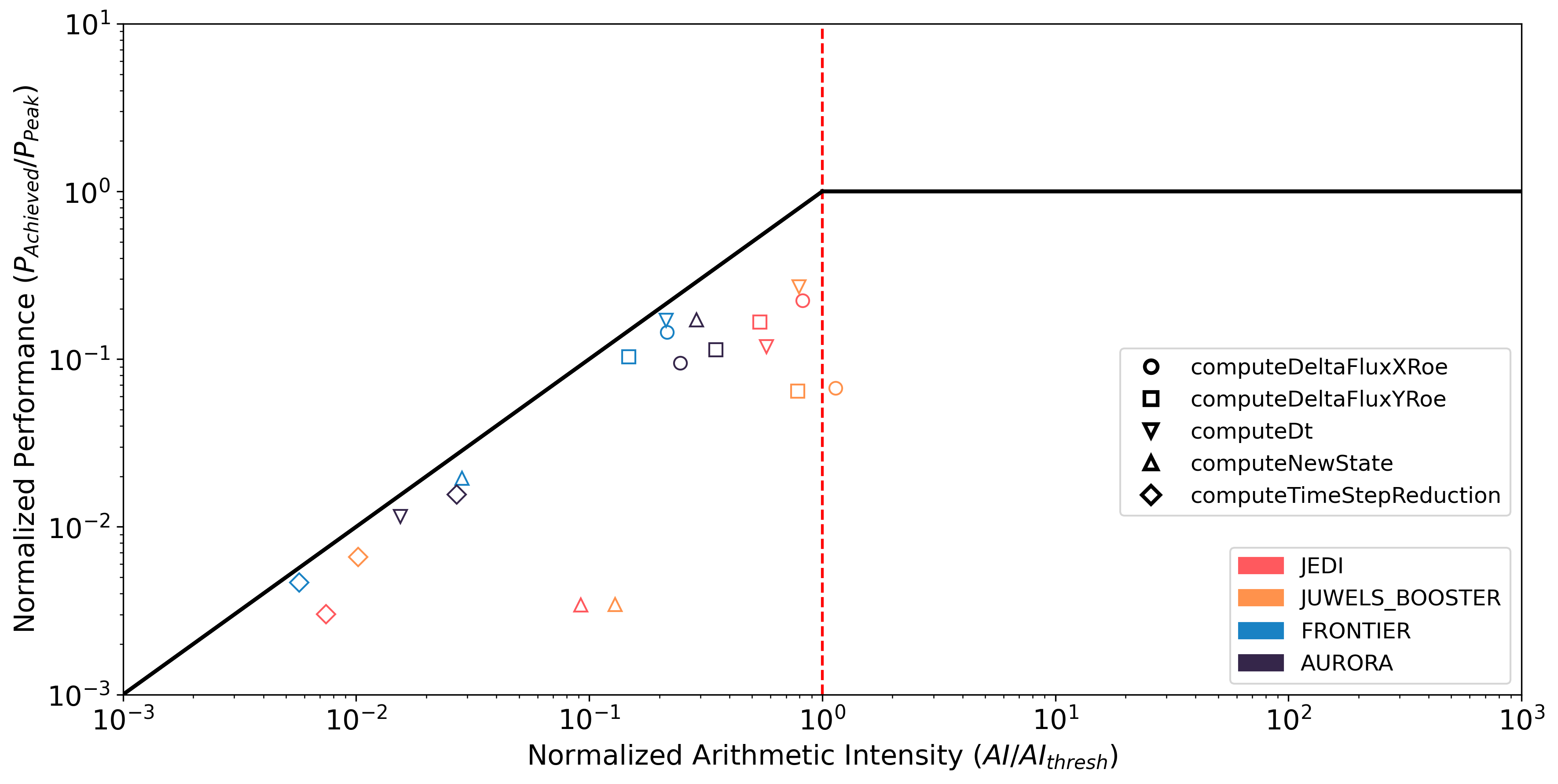}
        \caption{\centering Roofline graph for a circular dam break case with $N_\text{side}=15000$ for all architectures.}
        \label{fig:rooflineuni}
    \end{subfigure}\hfill
    \begin{subfigure}[t]{0.475\textwidth}
        \centering
        \includegraphics[width=0.975\linewidth]{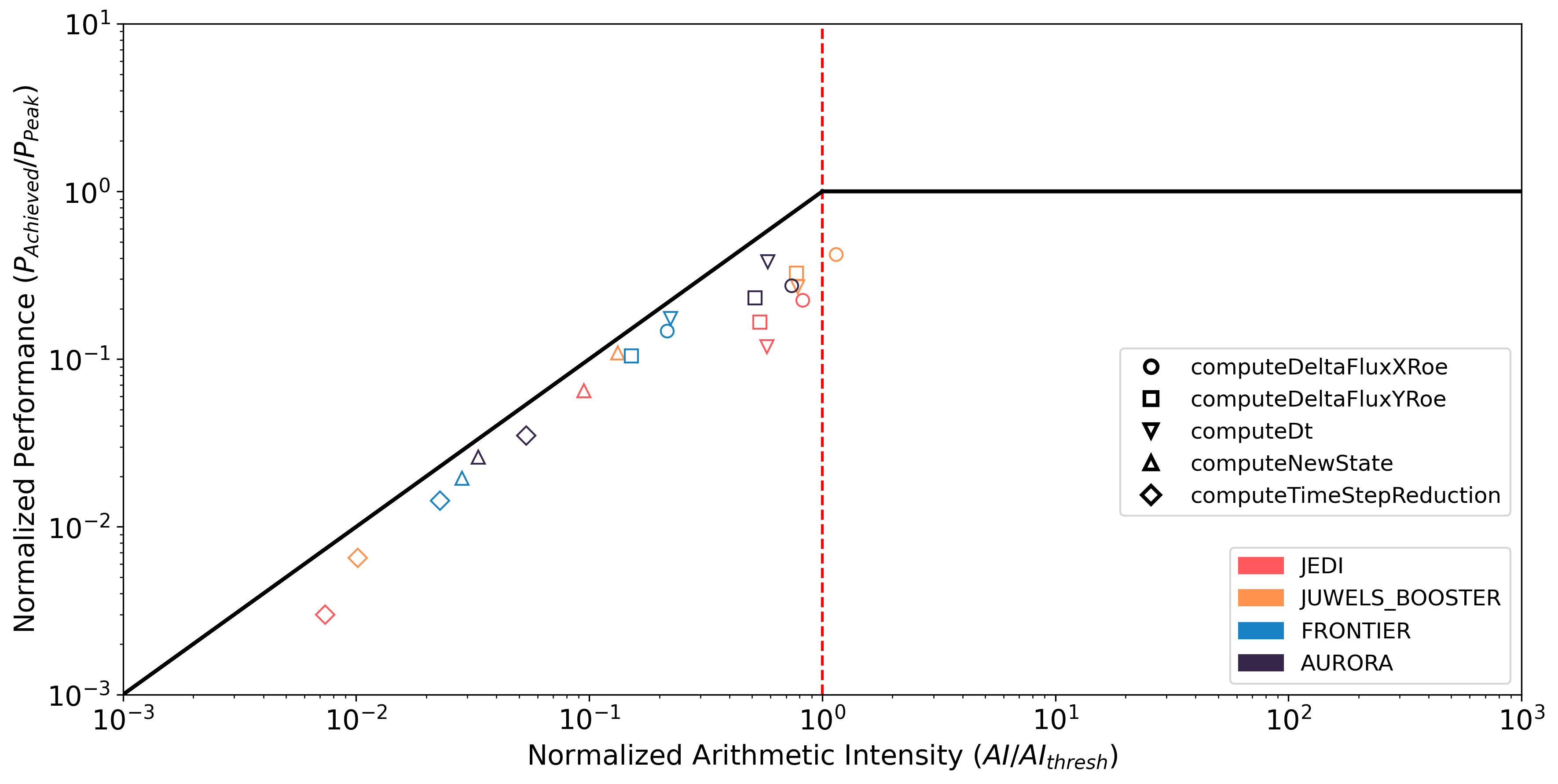}
        \caption{\centering Roofline for the best performing configuration of all architectures among problem sizes $N_\text{side}$ ranging from 4000 to 20000.}
        \label{fig:rooflinebest}
    \end{subfigure}
    \caption{\centering Roofline graphs for the five most intensive kernels in the circular dam break case.}
    \label{fig:roofline}
\end{figure*}

The plots in Figure \ref{fig:roofline} provide insights into the computational characteristics of the SERGHEI-SWE code's most intensive kernels during the circular dam break case. In both panels, we observe that the majority of the profiled kernels are situated within the memory-bound region of the normalized roofline, indicating that their performance is primarily limited by memory bandwidth rather than computational throughput. This points to the current flat memory access patters implemented in the loops showing up as inefficiencies or costly memory operations. This also highlights the need of remapping data to fast scratch memory.

In Figure \ref{fig:rooflineuni}, corresponding to a large fixed problem size ($N_\text{side} = 15000$), all architectures (JEDI, JUWELS BOOSTER, FRONTIER, and AURORA) exhibit a consistent pattern: key kernels such as \texttt{computeDeltaFluxXRoe}, \texttt{computeNewState}, and \texttt{computeDt} cluster along the sloped region below the bandwidth roofline. This suggests that despite differences in hardware architecture and peak compute capabilities, the kernels are similarly constrained by data movement. A significant shift toward compute-bound behavior is visible for the NVIDIA architectures, such as JEDI (GH200) JUWELS BOOSTER (A100). This is rather interesting given that most kernels are memory bound. This suggests a better handling of Kokkos kernels with the CUDA backend. 

Figure \ref{fig:rooflinebest} presents the best-performing configurations across a range of problem sizes ($N_\text{side}$ from 4000 to 20000). Here, the arithmetic intensity of some kernels slightly increases compared to the fixed problem size case, particularly for JEDI and FRONTIER, moving a few kernels closer to the bandwidth roof. This shift suggests that larger problem sizes can marginally improve computational efficiency. Nonetheless, even in these best cases, the operations remain fundamentally memory-bound.

An interesting observation from both plots is that \texttt{computeTimeStepReduction} consistently exhibits the lowest arithmetic intensity across all platforms, positioning it furthest from the ideal roof. This behavior highlights it as a potential target for further optimization, perhaps through improved memory access patterns, kernel fusion, or algorithmic restructuring to increase arithmetic intensity. However, this kernel typically demands only around 5\% of the simulation time.


\subsection{Performance Portability Analysis}

\begin{figure*}
        \centering
        \begin{subfigure}[t]{0.475\textwidth}
            \centering
            \begin{overpic}[width=\textwidth]{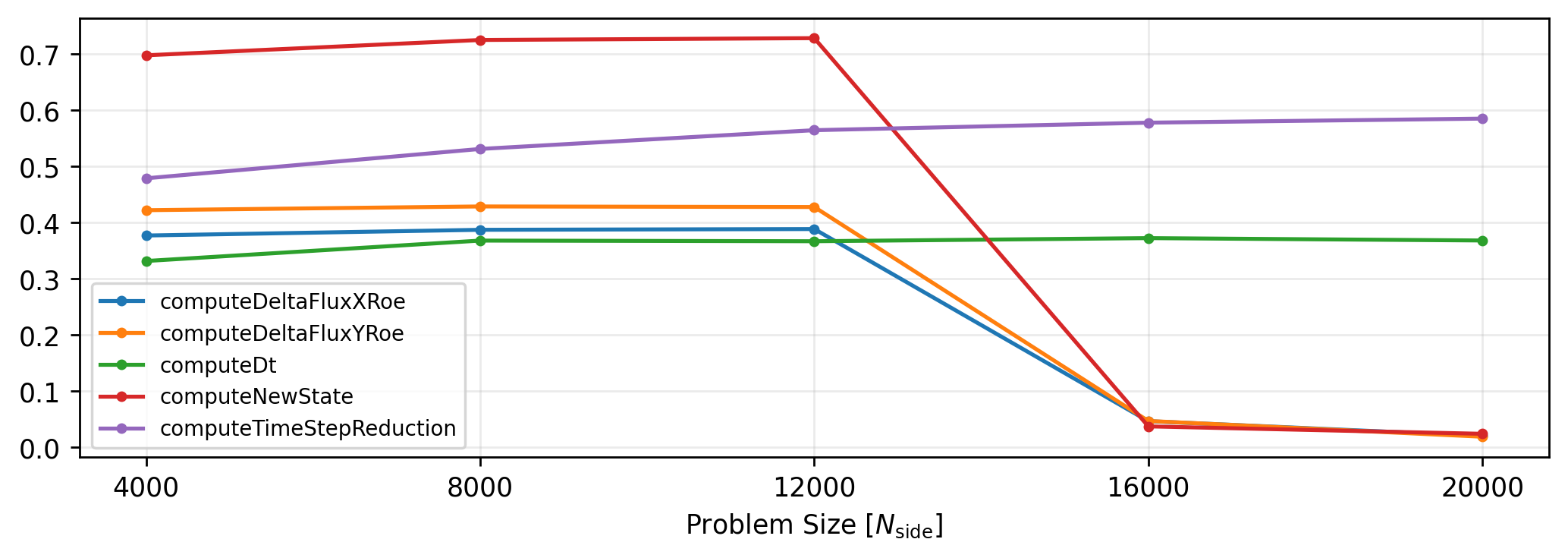}
                \put(0,37.5){\scalebox{0.8}{$\ppmetric_1$}}
            \end{overpic}
            \caption{\centering Variation of $\ppmetric_1$ with respect to problem size.}
            \label{fig:pp1}
        \end{subfigure}\hfill
        \begin{subfigure}[t]{0.475\textwidth}
            \centering
            \begin{overpic}[width=\textwidth]{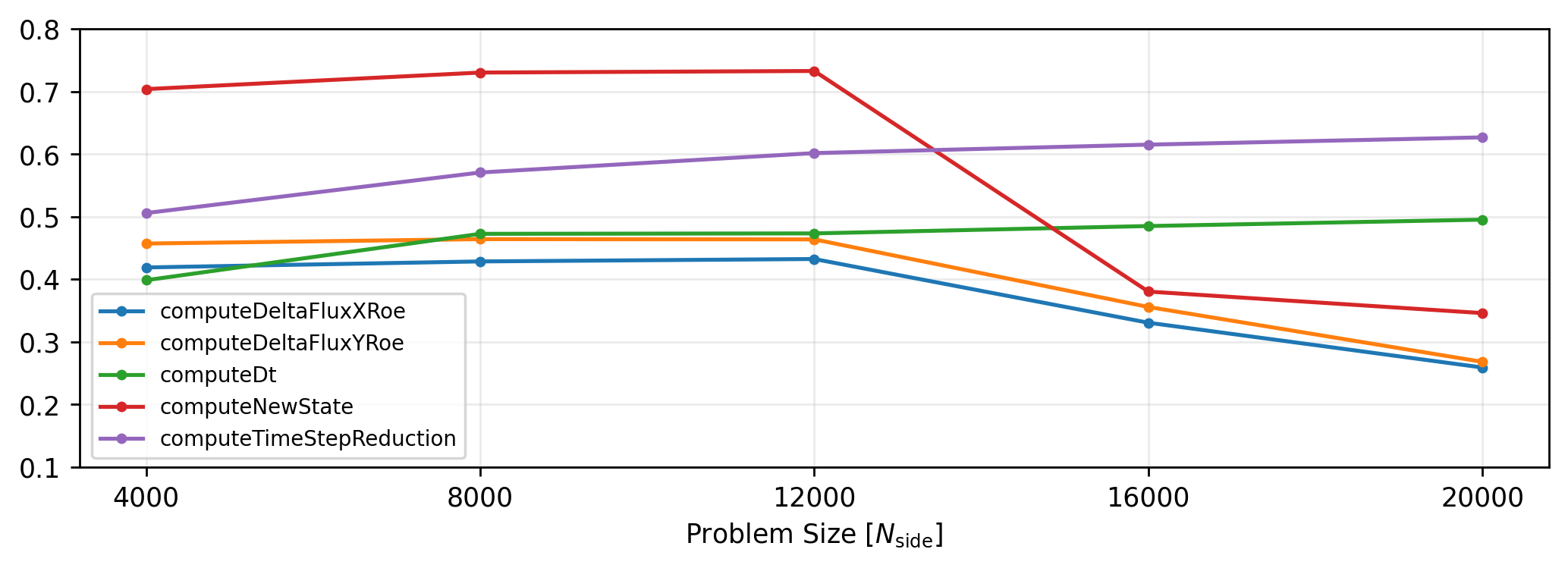}
                \put(0,37.5){\scalebox{0.8}{$\ppmetric_2$}}
            \end{overpic}
            \caption{\centering Variation of $\ppmetric_2$ with respect to problem size.}
            \label{fig:pp2}
        \end{subfigure}\hfill
        \caption{\centering Variation of performance portability metrics with respect to problem size. We observe that for $\protect\ppmetric_1$, the relative performance exhibits abrupt changes, similar to the raw performance shown in Figure~\ref{subfig:nxvariation}. In contrast, $\protect\ppmetric_2$ does not penalize performance drops as severely.}
        \label{fig:pp}
\end{figure*}

\begin{table}
    \centering
    \caption{\centering Performance portability metric values for a fixed problem size at $N_\text{side} = 15000$ }
    \begin{tabular}{l|cc}
    \hline
    \multirow{2}{*}{Kernel Name} & \multicolumn{2}{c}{Uniform Test} \\
     & $\ppmetric_1$ & $\ppmetric_2$ \\
    \hline
    \texttt{computeDeltaFluxXRoe}       & 0.1779 &  0.3350 \\
    \texttt{computeDeltaFluxYRoe}       & 0.2090 &  0.3689 \\
    \texttt{computeDt}                  & 0.3722 &  0.4847 \\
    \texttt{computeNewState}            & 0.0596 &  0.3736 \\
    \texttt{computeTimeStepReduction}   & 0.5731 &  0.6111 \\
    \hline    
    \end{tabular}
    \label{tab:portability}
\end{table}
 Table~\ref{tab:portability} presents the performance portability metrics $\ppmetric_1$ and $\ppmetric_2$ for the five most intensive kernels under a fixed problem size. The \texttt{computeTimeStepReduction} kernel  achieves the highest performance portability with $\ppmetric_1 = 0.5731$ and $\ppmetric_2 = 0.6111$, suggesting consistent and efficient performance across architectures even without fine-tuning. However, as we mentioned before this kernel may be fused with other kernels as it does not play a critical part of our simulation. In contrast, \texttt{computeNewState} shows the lowest portability ($\ppmetric_1 = 0.0596$), indicating $\ppmetric_1$ presents a high sensitivity to relative performance differences. Figure \ref{fig:pp} shines a light on the differences of each metric. 

Comparing the two metrics, $\ppmetric_2$ (arithmetic mean) values are consistently higher than $\ppmetric_1$ (harmonic mean) across all kernels and tests. This is expected, as the harmonic mean is more sensitive to low-performing outliers, thus providing a more conservative estimate of portability. Even if the validity of $\ppmetric_1$ has been heavily discussed with respect to the definition of performance portability \cite{Marowka2021reform} it might prove useful to assess portability from a more rigorous point of view, perhaps in a way to measure architecture-specific optimization for performance portable codes.

Regarding the SERGHEI-SWE code, we observe that its portability remains mostly under 70\%, indicating potential areas for optimization. These optimizations can again be guided by this same performance portability analysis, focusing on bottlenecks such as memory access patterns, kernel launch overhead and low occupancy.

    Establishing a fair experimental setup across GPU architectures proved difficult, underscoring the challenges of evaluating performance portability using architectural efficiency. Many hardware dependent variables such as memory bandwidth, compute throughput, and memory hierarchy strongly influence performance. Our Kokkos-based code, while portable, lacks tunable parameters to adapt to each architecture’s strengths. For example, setting fixed problem sizes for all architectures leads to resource under-utilization in some cases. In other cases, setting problem sizes based on a uniform fraction of available GPU memory may lead to mismatched workloads: systems with more memory processed larger problems, distorting comparisons. This revealed a broader issue: there are few clear, singular points of comparison across heterogeneous platforms. These difficulties highlight the importance of careful benchmarking and the need for software that is not only portable, but fairly comparable across systems.

    Performance portability analysis from an application efficiency point of view seems to avoid many of these issues, as it relies on comparing each architecture’s performance against its own optimized baseline. Rather than enforcing a uniform problem size or relying on theoretical peak values, this approach evaluates how efficiently a portable application utilizes the available hardware. By comparing the performance of a portable version of the code to an architecture-specific optimized version, we can assess how much performance is retained without architecture-specific tuning. This method reduces the bias introduced by architectural asymmetries and avoids misleading conclusions drawn from raw runtime or peak-normalized metrics. 
    
    However, for codes developed using performance portable frameworks, such as the SERGHEI-SWE solver, this approach incurs a considerable amount of additional effort. Constructing and maintaining highly optimized, architecture-specific versions purely for benchmarking defeats the purpose of portability and diverts resources away from scientific or engineering goals. As a result, while application efficiency comparisons offer a rigorous perspective, they may not be practical or sustainable in real-world research and development workflows, especially when the goal is to evaluate how well a single implementation performs across platforms.
    
 
    In this direction, developing a clear and fair methodology for evaluating performance portability across architectures could provide a consistent benchmark for the ever growing body of performance-portable software. Such an approach would help avoid diverting valuable resources from critical scientific efforts, such as developing more accurate flood simulation models and their optimization.


\section{Final Remarks}\label{sec:final}

\subsection{Conclusions}
We have conducted a performance analysis of the SERGHEI-SWE solver on four different HPC systems featuring four GPU architectures from the top vendors (NVIDIA, AMD, Intel) focusing on scaling performance, kernel performance and performance portability aspects of code. 

Our strong scaling results show efficient parallel performance up to 512--1024 GPUs across all architectures, with JEDI’s NVIDIA GH200 system achieving the fastest execution times, closely followed by Frontier's AMD MI250X platform. Aurora’s Intel Max 1550 and JUWELS Booster’s NVIDIA A100 showed comparable execution times, indicating similar levels of scalability and efficiency despite architectural differences. Weak scaling analysis further confirms the solver's robustness, maintaining high scalability (>90\%) as the problem size grows proportionally to the computational resources. 

The roofline analysis reveals that the most computationally intensive kernels are predominantly memory-bound across all platforms, highlighting memory bandwidth as a critical factor for achieving high performance. Additionally, variations in problem size show that the arithmetic intensity and performance of several kernels change in peculiar ways, particularly on NVIDIA architectures. We attribute this behavior to complex function calls within kernels, which may lead to register spilling and reduced occupancy on the device.

The performance portability study was based on both harmonic and arithmetic mean-based metrics demonstrated the portability of the SERGHEI-SWE code to be less than 70\% for all kernels. This fact suggests that this solver can benefit from optimizations regarding memory layout and access patters specifically increasing granularity in memory management. 

Performance portability metrics were based on architectural efficiency which directly measures performance consistency with respect to attainable peak hardware performance. This has the benefit of not needing a custom implementation for every architecture but the drawback of not taking into account attainable peak algorithm performance which is provided by highly optimized architecture specific implementations. 

These insights can help users of the SERGHEI-SWE code characterize their runtime depending on the GPU architectures they implement for their research. It also gives key information about the current state of the solver that they can take into account when migrating from one family of GPUs to another. 

Overall, while we see excellent behavior of the SERGHEI-SWE solver in terms of scalability. GPU kernel execution and performance portability analyses shine a light into device behavior which we can optimize. Doing this can further improve scalability and make the SERGHEI-SWE code an even better alternative for EWS and their applications. 

\subsection{Future Work}
    Future research with the SERGHEI-SWE needs to involve optimization of GPU kernels to tackle certain problems found in this study such as implementing more granular memory management using the Kokkos team concepts and the implementation of tunable architecture specific parameters such that parameter optimization can be achieved either manually or by implementing new tools such as Kokkos autotuning \cite{kokkos2024autotuning}. In the same direction, instrumentation using Kokkos tools might prove to be an advantage for Kokkos related debugging and testing. 

    As the SERGHEI-SWE solver is part of a bigger environment of SWE-related software, this work can be used as a methodology for benchmarking their performance and analyzing where optimization is needed. This is a critical task for EWS related software not only in hydrology but in numerical forecasting overall, hence it can also be an initial methodology to benchmark future performance portable numerical models.

    Establishing a fair process for comparison using architectural efficiency is also within future research, as performance portability is becoming more relevant to leverage the current exascale era.

    
\section*{Acknowledgment}

This research used resources of the Argonne Leadership Computing Facility, a U.S. Department of Energy (DOE) Office of Science user facility at Argonne National Laboratory and is based on research supported by the U.S. DOE Office of Science-Advanced Scientific Computing Research Program, under Contract No. DE-AC02-06CH11357.

This research also used resources of the Oak Ridge Leadership Computing Facility at the Oak Ridge National Laboratory under project GEO161: Advancing flood modeling with the SERGHEI code, which is supported by the Office of Science of the U.S. Department of Energy under Contract No. DE-AC05-00OR22725.  

We would like to acknowledge gratefully the Earth System Modelling Project (ESM) for supporting this work by providing computing time on the ESM partition of the JUWELS supercomputer at the Jülich Supercomputing Centre (JSC) through compute time projects RUGSHAS (project number 29000) and EHRTAS (project number 59866). 

The authors would also like to acknowledge the JUPITER Research and Early Access Program (JUREAP) for providing access to JEDI, the JUPITER Exascale Development Instrument.

\section*{Reproducibility}
    Scripts are provided for compiling the code, and reproducing the results presented in this study can be found in \cite{serghei_perfport_tests,serghei_gitlab}.
\printbibliography



%

\end{document}